\documentclass[runningheads]{llncs}

\usepackage[english]{babel}
\usepackage{blindtext}

\usepackage{booktabs}
\usepackage{multirow}
\usepackage{graphicx}
\usepackage{subcaption}
\usepackage[table,xcdraw]{xcolor}
\usepackage{enumitem}
\usepackage{cite}
\usepackage{siunitx}

\usepackage{cleveref}

\usepackage[absolute,showboxes]{textpos}

\usepackage{algorithm}
\usepackage[noend]{algpseudocode}
\usepackage{float}

\newfloat{algorithm}{t}{lop}

\newcommand{\lastdate}{August 2022\xspace}
\newcommand{\lastdatefull}{August 31\(^{st}\), 2022\xspace}
\newcommand{\godaddydrop}{20.3\%\xspace}
\newcommand{\godaddydropabs}{45.6\,M\xspace}
\newcommand{\vsixendrespops}{10\%\xspace}
\newcommand{\vsixendtoptenzoneshare}{97.5\%\xspace}
\newcommand{\vsixendtoptenabszoneshare}{24.8\%\xspace}
\newcommand{\vsixresstart}{11.4\%\xspace}
\newcommand{\vsixresend}{55.1\%\xspace}
\newcommand{\vsixchgjansev}{17.3\%\xspace}
\newcommand{\alexaonevsixresstart}{38.9\%\xspace}
\newcommand{\alexaonevsixresend}{80.6\%\xspace}

\newcommand{\lastunresvsix}{44.9\%\xspace}
\usepackage{colortbl}
\usepackage{color}

\definecolor{atomictangerine}{rgb}{1.0, 0.6, 0.4}
\definecolor{RED}{rgb}{0.8, 0.0, 0.0}

\newcommand{\phantomsubfigure}[1]{%
    \begin{subfigure}[t]{0pt}%
        \phantomcaption%
        \label{#1}%
    \end{subfigure}%
}

\newcommand{\notext}[1]{}
\newcommand{\ie}{i.e., \@}
\newcommand{\eg}{e.g., \@}

\newcommand{\anss}{authoritative nameservers\xspace}
\newcommand{\ns}{nameserver\xspace}
\newcommand{\nss}{nameservers\xspace}

\newcommand{\nsset}{NS set\xspace}
\newcommand{\nssets}{NS sets\xspace}

\newcommand{\broken}{broken IPv6-delegation\xspace}
\newcommand{\Broken}{Broken IPv6-delegation\xspace}

\newcommand{\ld}{lame delegation\xspace}

\newcommand{\glue}{GLUE\xspace}
\newcommand{\GLUE}{GLUE\xspace}

\newcommand{\NS}{\texttt{NS}\xspace}
\newcommand{\NSes}{\texttt{NS}es\xspace}
\newcommand{\A}{\texttt{A}\xspace}
\newcommand{\AAAA}{\texttt{AAAA}\xspace}

\newcommand{\resip}[1]{IPv{#1}-resolvable\xspace}
\newcommand{\ip}[1]{IPv{#1}\xspace}

\newcommand{\ipo}[1]{IPv{#1}-only\xspace}

\begin{document}

\setlength{\TPHorizModule}{\paperwidth}
\setlength{\TPVertModule}{\paperheight}
\TPMargin{5pt}
\begin{textblock}{0.8}(0.1,0.02)
     \noindent
     \footnotesize
     If you cite this paper, please use the PAM reference:
     Florian Streibelt, Patrick Sattler, Franziska Lichtblau, Carlos H. Ga\~n\'an, Anja Feldmann, Oliver Gasser, and Tobias Fiebig.
     How Ready Is DNS for an IPv6-Only World?
     In \textit{Proceedings of the Passive and Active Measurement Conference 2023 (PAM ’23), March 21--23, 2023}.
\end{textblock}
 
\title{How Ready Is DNS for an IPv6-Only World?}

\author{Florian Streibelt\inst{1} \and
Patrick Sattler\inst{2} \and
Franziska Lichtblau \inst{1} \and
Carlos H. Ga\~n\'an\inst{3} \and
Anja Feldmann \inst{1} \and
Oliver Gasser\inst{1} \and
Tobias Fiebig\inst{1}}

\authorrunning{Streibelt et al.}

\institute{Max Planck Institute for Informatics
\email{\{fstreibelt,rhalina,anja,oliver.gasser,tfiebig\}@mpi-inf.mpg.de} \and
TU M\"unchen
\email{sattler@net.in.tum.de} \and
TU Delft
\email{c.hernandezganan@tudelft.nl} }

\maketitle

\begin{abstract}
DNS is one of the core building blocks of the Internet.  In this paper, we
investigate DNS resolution in a strict \ipo6 scenario and find that a
substantial fraction of zones cannot be resolved.  We point out, that the presence of an
\AAAA resource record for a zone's \ns does not necessarily imply that it is
resolvable in an \ipo6 environment since the full DNS delegation chain must resolve via \ip6 as well.
Hence, in an \ipo6 setting zones may experience an effect similar to what is commonly referred to as \ld.

Our longitudinal study shows that the continuing centralization of the Internet
has a large impact on \ip6 readiness, \ie a small number of large DNS providers
has, and still can, influence \ip6 readiness for a large number of zones.  A
single operator that enabled \ip6 DNS resolution--by adding \ip6 glue
records--was responsible for around \godaddydrop of all zones in our
dataset not resolving over \ip6 until January 2017.  Even today, \vsixendrespops of DNS operators are responsible for more
than \vsixendtoptenzoneshare of all zones that do not resolve using \ip6.
\end{abstract}
\section{Introduction}

With the recent exhaustion of the IPv4 address space, the question of IPv6
adoption is gaining importance. More end-users are getting IPv6 prefixes from
their ISPs, more websites are reachable via IPv6, hosting companies start billing
for IPv4 connectivity or give discounts for \ipo6 hosting and IoT devices further push
IPv6 deployment. Yet, one of the main entry-points for Internet services---the
DNS---is suffering from a lack of pervasive IPv6 readiness.  While protocols
such as Happy Eyeballs \cite{rfc6555,rfc8305} help to hide \ip6 problems, they
complicate detection and debugging of IPv6 issues.  Indeed, the threat of DNS
name space fragmentation due to insufficient \ip6 support was already predicted
in RFC3901, over 18 years ago~\cite{rfc3901}. Hence, in this paper, we
measure the current state of IPv6 resolvability in an \ipo6 scenario.

\begin{figure}[t]
\centering
\includegraphics[keepaspectratio,width=0.99\linewidth]{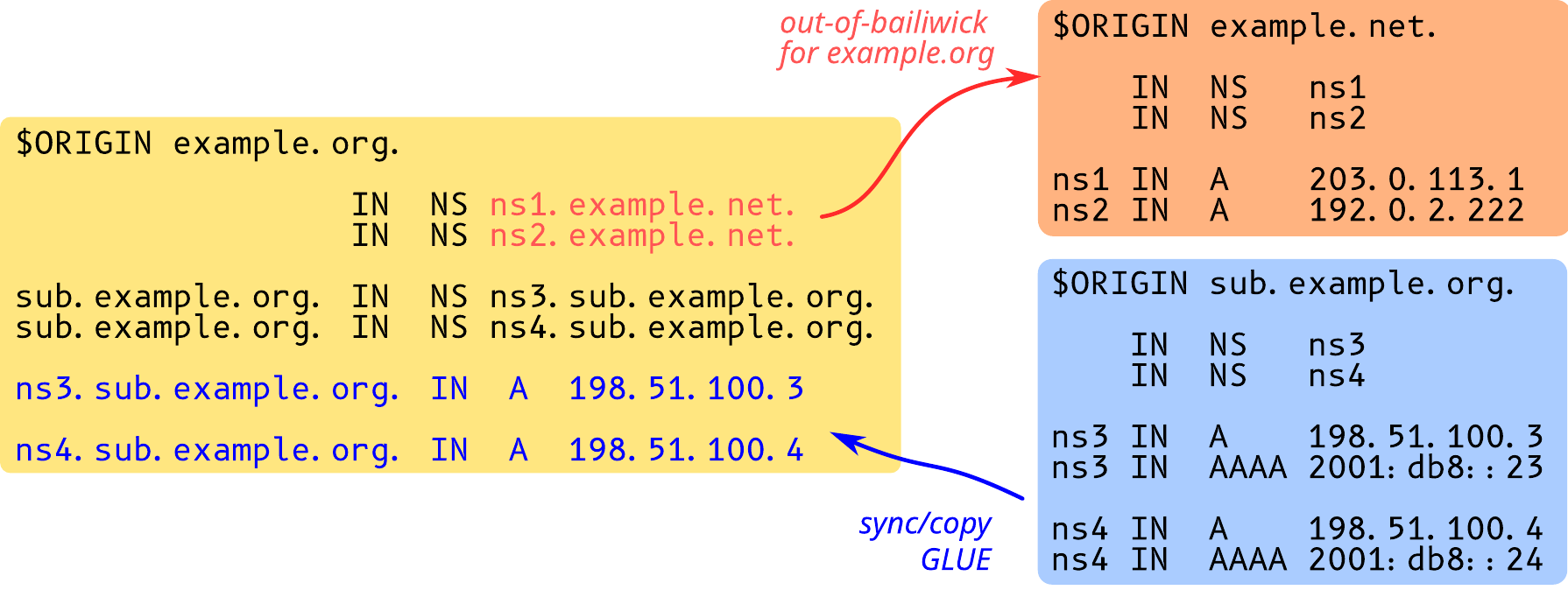}
\caption{\Broken for \texttt{example.org} (missing \AAAA resource records in
\texttt{example.net} for \NS) and \texttt{sub.example.org} (missing \ip6-\GLUE in parent).}
\label{fig:brokenv6dns}
\end{figure}

In \Cref{fig:brokenv6dns} we show two common misconfigurations, which prevent DNS
resolution over \ip6 and lead to an effect similar to what is commonly called \ld.
Note, that RFC8499\cite{rfc8499} defines \ld as incorrect \NS entries or \nss
\emph{not responding properly}. While the observed behaviour might look the same,
the underlying misconfiguration, \eg missing \AAAA or \glue for \ip6, often is different.
Hence, in this paper we use the term \broken to avoid unnecessary ambiguity and distinguish
the case of zones that are not \ip6 ready, e.g. show no intent to support \ip6 by not having any \AAAA
records, and zones that appear to intend supporting \ip6, Section~\ref{sec:background}.

In the first example, the external \nss (``out-of-bailiwick'') of
\texttt{ex\-am\-ple.org} do not have \AAAA records and, thus, the resolution
via \ip6 is impossible. In the second example, the zone
\texttt{ex\-am\-ple.org} misses the \AAAA glue records.
These glue records make the \A/\AAAA records available for resolution if they 
have to be resolved from the zone being delegated, i.e., the names of the \NS
\texttt{\{ns3,ns4\}.sub.ex\-am\-ple.org} are in-bailiwick.

These examples highlight (a) that it needs cooperation between multiple parties
for proper configuration, \ie \texttt{sub.ex\-am\-ple.net} cannot be resolved
via \ip6 even though it is correctly configured; (b) that dual-stack hides
issues, \ie both examples work for dual-stack enabled hosts where the \AAAA
records for \texttt{ns3} and \texttt{ns4} are resolvable. This demonstrates
how working IPv4 resolution hides \broken for dual-stack DNS recursors.

To be \ip6 \emph{ready}, DNS resolution must work in \ipo6 scenarios. 
In this paper, we leverage passive DNS data---the Farsight SIE
dataset~\cite{Farsight-SIE}---to identify scenarios in which the DNS delegation
chain breaks when only IPv6 is available. Our main contributions can be
summarized as follows:

\begin{itemize}[leftmargin=*]
\item We identify common \broken scenarios and point out the importance of
 checking the full delegation chain.

\item We show that big players have a major impact on the number of zones
 affected by \broken. Today, 10 DNS providers are responsible for about \vsixendtoptenabszoneshare of
 \ipo6-unresolvable domains we observe. Just by adding correct glue records,
 in Jan.\ 2017 one single provider fixed the \ipo6 name resolution
 of more than \godaddydropabs domains (\godaddydrop of the domains in the dataset).

\item Resilience mechanisms often hide misconfigurations. For example,
 \broken is hidden by the combined efforts of DNS resilience and Happy
 Eyeballs. Correctly configuring ones own DNS zone is not
 sufficient and dependencies are often non-obvious.

\item Additionally, we conduct a thorough validation of our methodology.
We assess the coverage of the Farsight SIE data in comparison to available ground-truth zonefile data, finding it to provide sufficient coverage for our analysis.
Furthermore, we cross-validate our passive measurement results using active measurements, again finding our results to be robust.

\item We implemented a DNS measurement tool instead of using, e.g., ZDNS~\cite{izhikevich2022zdns}, as we need IPv6 support which ZDNS does not (yet) support.
The dataset from our active measurements and an implementation of our scanning methodology, including a single-domain version operators can use to evaluate IPv6 support for their own domains, are publicly available at:\\ \url{https://github.com/mutax/dns-v6-readyness}
\end{itemize}

\section{Broken IPv6 Zone Delegation}
\label{sec:background}

In this section, we briefly recap DNS zone delegation, and sketch common DNS
resolution failure scenarios.

\subsection{Background: DNS Zone Delegation}

The DNS is organized in a hierarchical structure where each node represents a
zone that can be operated separately from its parent or child zones.
For a zone to be resolvable, \NS resource records have to be set in two places.
First, the parent of the zone has to explicitly delegate the zone to
\anss via \NS resource records. If an authoritative
server has a domain name within the delegated zone itself or a child zone, i.e., if it is
``in-bailiwick''~\cite{rfc8499}, the parent zone must also contain \A and \AAAA resource
records for this name, called \GLUE, that are returned in the
\texttt{ADDITIONAL} section of the DNS responses whenever the \NS record is
returned. This process breaks the circular dependency in the resolution chain.
Furthermore, the zone itself must contain appropriate \NS records as well as \A
and \AAAA records if they are in-bailiwick.
If the name in an \NS record is not within the zone itself or a child zone, i.e., it is
out-of-bailiwick, then the zone of the \NS' name must also resolve
for the initial zone to be resolvable.

\subsection{Reasons for Broken \ip6 Delegation}

In this paper, we focus on a subset of DNS misconfigurations. In an \ipo6 scenario these
misconfigurations can lead to effects similar to what is commonly referred to as \ld.
To avoid ambiguity, we use the term \broken referring to any set of misconfiguration
specific to \ip6, that breaks the DNS delegation chain of a zone and prevents any of
its records from resolving in an \ipo6 scenario.
Other issues where a zone does not resolve due to, e.g., DNSSEC problems or unresponsive
\nss, i.e., the strict definition of ``lame delegation'' (see RFC8499~\cite{rfc8499})
are out-of-scope.
The issues we discuss can also occur in \ip4 DNS resolution, but are usually quickly
discovered given the currently still large number of sites with \ip4-only connection to the Internet,
that will not be able to resolve the affected zones.

For a zone to be {\resip6} ---\ie resolvable using \ipo6--- the zones of the
\anss have to be resolvable via \ip6 and at least one \ns must be
accessible via \ip6. This has to be the case \emph{recursively}, \ie
not only for all parents of the zone itself but also for all parents of the
\anss in such a way that at least for one\footnote{RFC2182~\cite{rfc2182} suggests to avoid such single points of failure} of the authoritative nameservers of
a zone a delegation chain from the root zone exists, that is fully resolvable using \ip6.
We identify the following misconfigurations which can cause \broken in an \ipo6 setting:

\begin{itemize}[noitemsep,topsep=0pt,parsep=0pt,partopsep=0pt,leftmargin=*] %

\item \textbf{No \AAAA records for \NS names:} If none of the \NS records for a
  zone in their parent zone have associated \AAAA records, resolution
  via \ip6 is not possible.

\item \textbf{Missing \GLUE:} If the name from an \NS record for a zone is in-bailiwick,
  \ie the name is within the zone or below~\cite{rfc8499}, a parent zone must
  contain an \ip6 \GLUE record, i.e., a parent must serve the corresponding
  \AAAA record(s) as \texttt{ADDITIONAL} data when returning the \NS record in
  the \texttt{ANSWER} section.

\item \textbf{No \AAAA record for in-bailiwick \NS:} If an \NS record of a
  zone points to a name that is in-bailiwick but the name lacks \AAAA record(s)
  in its zone, \ipo6 resolution will fail even if the parent provides \GLUE, when the
  recursive server validates the delegation path. One such example is
  Unbound~\cite{unbound} with the setting \texttt{harden-glue: yes}--the default.

\item \textbf{Zone of out-of-bailiwick \NSes not resolving:} If an \NS record
  of a zone is out-of-bailiwick, the corresponding zone must be \resip6
  as well. It is insufficient if the name pointed to by the \NS record has an
  associated \AAAA record.

\item \textbf{Parent zone not \resip6:} For a zone to be resolvable via
  \ip6 the parent zones up to the root zone must be \resip6. Any non-\resip6
  zone breaks the delegation chain for all its children.

\end{itemize}

The above misconfigurations are not mutually exclusive. For example, if the \NS
sets between parent and child differ, a common misconfiguration~\cite{sommese},
the \NS in the parent may not resolve due to missing \GLUE (as they are
in-bailiwick) \emph{but also} the \NS in the child may not resolve due to having no
\AAAA for their names, if they are out-of-bailiwick.
In this paper we investigate the prevalence of these misconfigurations to
evaluate the IPv6 readiness of the DNS ecosystem.

\section{Datasets and Methodology}
In this section, we present our choice of datasets as well as our active and passive measurement methodology
for identifying DNS misconfigurations that break \ipo6 resolution.

\subsection{DNS Dataset: Farsight SIE}
\label{sec:method:dataset}

For our evaluation we are looking for a dataset that enables us to (a) perform a
longitudinal study, (b) detect \ip6 DNS misconfigurations, (c)
analyze not just top level domains (TLDs) but \emph{also} zones deeper in the
tree, and (d) focus on zones that are used in-the-wild. As such we select the
Farsight SIE dataset for our study.

The \textit{Farsight Security Information Exchange} (SIE)
dataset~\cite{Farsight-SIE} is collected by Farsight Inc.\ via globally
distributed sensors, co-located with recursive DNS resolvers. Each sensor
collects and aggregates all DNS cache misses that the recursive DNS resolver
encounters, i.e., the outgoing query and the received answer. By only
recording cache-misses and providing aggregates, Farsight reduces the risk
of exposing Personally Identifiable Information (PII).
Cache-misses occur when a recursive DNS resolver does not have a DNS record for a specific domain name in its cache (or the record's TTL has expired).
The recursive resolver then has to ask the authoritative nameserver for the requested name, which is then recorded by Farsight SIE.
Farsight does not share the exact number and location of its sensors for business confidentiality reasons.
Farsight's SIE dataset has been used in previous research \cite{fiebig2017something,liu2018reexamination,houser2021comprehensive} and its efficacy, coverage, and applicability for research has been demonstrated in the past~\cite{foremski2019dns}.
We discuss ethical considerations of using this dataset in \Cref{sec:ethics}.

We use monthly aggregates from January 2015 to \lastdate, containing unique
tuples of: requested name, requested RRtype, bailiwick of the response, and
returned data record, also for the additional sections, see
\Cref{tab:data}. Thus, the Farsight dataset contains essential
information for us, as it also records additional data as entries with the
bailiwick of the parent.
In addition, the Farsight dataset reaches deeper into the DNS hierarchy than,
e.g., OpenINTEL~\cite{openintel}, as it monitors DNS requests in the wild instead of
resolving a set of names below zones sourced from TLD zone files.

\begin{table}[t!]
\centering
\scriptsize
\caption{List of data fields in the Farsight SIE dataset.}
\label{tab:data}
\begin{tabular}{lp{.42\columnwidth}p{.3\columnwidth}}
\toprule
\textbf{Field} & \textbf{Description}                                                              & \textbf{Example}    \\ \midrule
count          & \# of times the tuple \texttt{<rrname, rrtype, bailiwick, rdata>} has been seen.  & \texttt{12}         \\
\rowcolor[HTML]{C0C0C0}
time\_first    & Unix timestamp of the first occurrence of the unique tuple during the data slice  & \texttt{1422251650} \\
time\_last     & Unix timestamp of the last occurrence of the unique tuple during the data slice.  & \texttt{1422251650} \\
\rowcolor[HTML]{C0C0C0}
rrname         & Requested name in the DNS.                                                        & \texttt{example.com}\\
rrtype         & Requested RRtype of the query.                                                    & \texttt{NS}         \\
\rowcolor[HTML]{C0C0C0}
bailiwick      & Zone authoritative for the reply.                                                 & \texttt{com}        \\
rdata          & List of all responses received in a single query.                                 & \texttt{{[}"ns1.example.com", "ns2.example.com"{]}} \\ \bottomrule
\end{tabular}
\end{table}

\begin{figure}[t!]
	\centering
	\includegraphics[width=\textwidth]{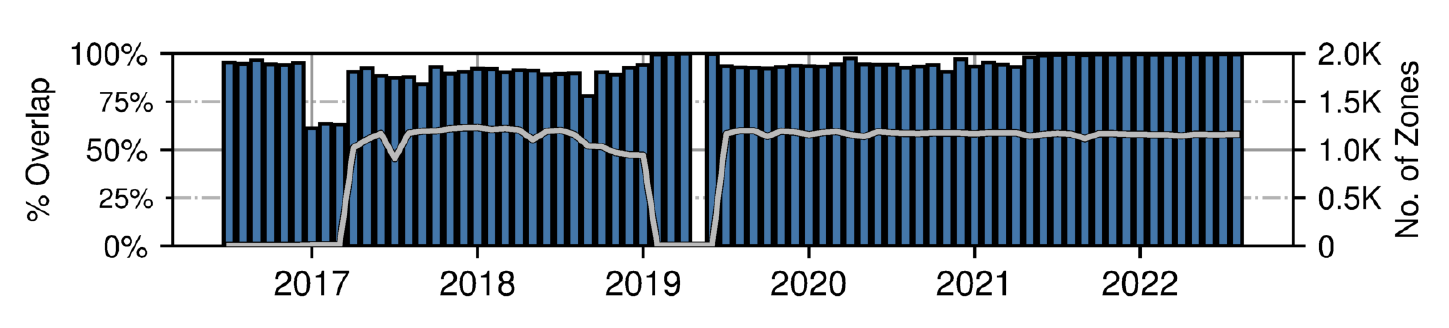}
	\caption{Zone coverage of Farsight data and number of zones used for the evaluation. We used available zone files to determine the share of covered second level domains by Farsight's dataset. 
	Please note the dip in the graph from February to August 2019, where our zone file collection was limited, i.e., we only collected few zones with high coverage (February - April and July, including .com), or no data at all (May and June).}
	\label{fig:farsighteval}
\end{figure}

\paragraph*{Farsight Global Zone Coverage}
A common question when using a passive dataset like the one provided by Farsight is how well it actually covers zones on the Internet.
In order to determine the coverage of the Farsight dataset, we evaluated the overlap of the second-level domains (SLDs) observed in the dataset with ground-truth data, i.e., the names extracted from available zone files.
Specifically, we are comparing to .com, .net, and other gTLD (generic TLD) zone files starting from mid of 2016.
Additionally, from April 2017 onward, we also obtained CZDS (ICANN Centralized Zone Data Service) zone file data for all available TLDs.
Moreover, we use publicly available zone file data from .se, .nu, and .ch for the coverage analysis.
In total, this allows us to compare Farsight's data to more than \num{1.1}k zones as of August 2022.

Looking at coverage over time, we find a significant overlap between the Farsight dataset and the number of actually delegated zones based on zone files, see \Cref{fig:farsighteval}.
Coverage averages above \SI{95}{\percent} from 2019 onwards, with especially since May 2021, our coverage reaches over \SI{99}{\percent}.
Furthermore, we find a reduced average coverage in the beginning of 2017.
A closer investigation revealed that these relate to the introduction of various vanity gTLDs with an overall small size, i.e., below 100 delegated zones in the TLD.
This implies that missing coverage for just a few zones would lead to a significant reduction in aggregate coverage.
Nevertheless, our analysis shows that a significant share of zones is covered in the Farsight dataset.
Hence, we the Farsight dataset--especially due to the historic perspective it provides--is ideal to investigate our research questions.

Despite this high coverage, we still face the drawback of the Farsight dataset relying on real-world usage. 
As such, a missing record in the passive dataset does not necessarily indicate non-existence.
Hence, we independently corroborate all major findings with data from TLD zone files for a specific period to check for missing glue records in the zone file, see \Cref{sec:discussion:limitations}.

\subsection{Domain Classification}
\label{sec:psl}

There are many ways to cluster DNS domains into subgroups. For example, one may
look only at the \emph{Top Level Domains} as specified by ICANN~\cite{tlds},
or use the \emph{Public Suffix List (PSL)} provided by the Mozilla
Foundation~\cite{psl} to identify second level domains. The PSL is used by
browser vendors to decide if a domain is under private or public control, e.g.,
to prevent websites from setting a \emph{super-cookie} for a \emph{domain} such
as \texttt{.co.uk}. Based on matching monthly samples of the ICANN TLDs and
the PSLs we identify \emph{TLDs} as well as \emph{2\(^{nd}\) Level Domains},
and \emph{Zones Below 2\(^{nd}\) Level}, i.e., all zones \emph{below}
2\(^{nd}\) Level Domains.

Another way of grouping DNS domains is to use the Alexa Top-1M
list~\cite{Alexa}. Using, again, matching monthly samples, we distinguish
between the Top~1K, Top~1K--10K, Top~10K--100K, and Top~100K--1M domains. We
note that there are limitations in the Alexa Top
List~\cite{scheitle2018long,rweyemamu2019clustering}, but compared to other
toplists such as Tranco~\cite{le2019tranco}, the Alexa list is available
throughout the measurement period.

\subsection{Misconfiguration Identification}

Here, we describe how we identify whether zones can be resolved only via IPv4,
only via IPv6, via IPv4 and IPv6, or not at all from the dataset.

\noindent{\bf 1.\ Per zone \nsset identification:}
We first identify all zone delegations by extracting all entries with
\texttt{rrtype} \(=\) \texttt{NS}. Next, for all names used in these
delegations, we find all associated IPs by extracting all \texttt{A} and
\texttt{AAAA} records. We do not consider
\texttt{CNAME}s since they are invalid for \NS entries, see
RFC2181~\cite{rfc2181}.

We then iterate over all zones, i.e., names that have \texttt{NS} records, to
create a unique zone list. In this process, we record the \NS records for each
bailiwick sending responses for this zone observed in the dataset, and for each
\NS name all \AAAA and \A type responses, again grouped by bailiwick from which
they were seen. This also captures cases where parent and child return
\emph{different} \NS sets.

\noindent{\bf 2.\ Per zone DNS resolution:}
We consider a zone to be resolvable via \ip4 or \ip6 if \emph{at least one} of the \NS listed for the zone can be resolved via \ip4 or \ip6 respectively.
Hence, to check which zones can be resolved using which IP protocol version we simulate the
DNS resolution, starting at the root, i.e, we assume the Root zone \texttt{.} to be resolvable by \ip4 and \ip6. 
We then iterate over the zone set with attached
\NS and \A/\AAAA data.
For each zone, except the root zone, we initialize an empty state marking the
zone as not resolving.

We then attempt to resolve each zone. For that, we first check if the zone's
parent has been seen.

If so we check for each \texttt{NS} of the zone we are trying to resolve as
listed in the parent whether its name resolves via IPv4 and/or IPv6. This is
the case if:
\begin{enumerate}
\item The \NS is outside the zone we are trying to resolve, the \texttt{NS}'
zone has been recorded as resolving in the zone state file (via IPv4 and/or
IPv6), and there are \A/\AAAA records with that zone's bailiwick for the \NS.
\item The \NS is in the zone we are resolving and there is an \A/\AAAA glue
record for the name with the bailiwick of the zone's parent (only if an
in-bailiwick \NS is listed in the parent).  \end{enumerate}

\algdef{SE}[DOWHILE]{Do}{doWhile}{\algorithmicdo}[1]{\algorithmicwhile\ #1}
\begin{algorithm}[H]
	\caption{Resolve Zones from Passive Data}\label{alg:res_chain}
	\begin{algorithmic}[1]
		\State ${zone\_res \gets \{\}}$
		\State ${ns\_res \gets \{\}}$
		\State ${prev\_res\_zones \gets -1}$
		\State ${cur\_res\_zones \gets 0}$
		\State
		\While {!$prev\_res\_zones$ == $cur\_res\_zones$}
			\State $prev\_res\_zones \gets cur\_res\_zones$
			\State $cur\_res\_zones \gets 0$
			\For {$zone$ in $input$}
				\If {$zone\_res[zone.parent][res]$}
					\State ${glue\_resolve \gets false}$
					\State ${zone\_resolve \gets false}$
					\For {$NS$ in $glue$}
						\If {$NS$ in $ns\_res$ $||$ ($NS$ in $zone$ $\&\&$ $zone.parent$ has $NS.ip$) $||$ ($zone\_res[ns\_zone][res]$ $\&\&$ $ns\_zone$ has $NS.ip$)}
								\If {$zone\_res[ns\_zone][res]$ $\&\&$ $ns\_zone$ has $NS.ip$}
									\State ${ns\_res[NS] \gets true}$
								\EndIf
								\State ${glue\_resolve \gets true}$
						\EndIf
					\EndFor
					\For {$NS$ in $zone$}
						\If {$NS$ in $ns\_res$ $||$ ($NS$ in $zone$ $\&\&$ $zone$ has $NS.ip$) $||$ ($zone\_res[ns\_zone][res]$ $\&\&$ $ns\_zone$ has $NS.ip$)}
								\If {$zone\_res[ns\_zone][res]$ $\&\&$ $ns\_zone$ has $NS.ip$}
									\State ${ns\_res[NS] \gets true}$
								\EndIf
								\State ${zone\_resolve \gets true}$
						\EndIf
					\EndFor
					\State {$zone\_res[zone][glue\_res] \gets glue\_resolve$}
					\State {$zone\_res[zone][zone\_res] \gets zone\_resolve$}
					\If {$glue\_resolve$ $\&\&$ $zone\_resolve$}
						\State {$zone\_res[zone][res] \gets true$}
						\State {$cur\_res\_zones \gets cur\_res\_zones$ $+$ $1$}
					\EndIf
				\EndIf
			\EndFor
		\EndWhile
	\end{algorithmic}
\end{algorithm}
\vspace{-1.5em}

To ensure full resolution, we also have to check that the \NS listed in the
child resolve. For \NS with names under the zone this is the case if the \NS
listed for this zone in the parent can be reached via \ip4/\ip6, see above, and
they have \A/\AAAA records with the bailiwick of the zone itself. For
out-of-bailiwick \NS, this is again the case if their own zone resolves and they
have \A/\AAAA records.

A single iteration of this process is not sufficient, as zones often rely on
out-of-bailiwick \NS. Hence, we continue iterating through the list of zones
until the number of unresolved zones no longer decreases. For a simplified
pseudo-code description, see \Cref{alg:res_chain}.

\vspace{-0.5em}
\subsection{Active Measurement Methodology}
\label{sec:validation}

To validate our passive measurement results, we implemented a resolver in python.
While, technically, Izhikevich et al. presented ZDNS, a tool for this purpose, ZDNS does not provide sufficient support for \ip6 resolution for our use-case.
Our measurement methodology follows essentially the same algorithm as our passive resolution.
For each zone, we start at the root, and iterate through the DNS tree.
From there, we query all authoritative nameservers recorded in the parent on each layer of the DNS hierarchy using \ip4 and \ip6 where possible for the \NS of that zonelayer.
Furthermore, we try to obtain any possibly available \GLUE (\A and \AAAA) for in-bailiwick \NS. 
For out-of-bailiwick \NS, we try to resolve the \NS, again starting from the root.
If there is an inconsistency between parent and child, i.e., if we discover additional \NS when querying the \NS listed in the parent, we also perform all queries for this layer against these, noting that they were only present in the child.

To limit the amount of queries sent to each server, our implementation follows the underlying principles of QNAME minimization as described in RFC7816~\cite{rfc7816}. 
By using the NS resource record type to query the parent zones we can directly infer zonecuts and store GLUE records from the additional section, if present. 
Note that RFC8020~\cite{rfc8020} is still not implemented by all nameservers, thus we cannot rely on NXDOMAIN answers to infer that no further zones exist below the queried zone. 
Our measurement tool will retry queries using TCP on truncation and disable EDNS when it receives a FORMERR from the upstream server. 

To further limit the number of queries sent, all responses, including error responses or timeouts, are cached.
We limit the number of retries (4) as well as the rate (20 second wait time) at which they are sent.
To further enrich the actively collected dataset, we query all authoritative nameservers of a zone for the NS, TXT, SOA and MX records of the given zone as well as the version of the used server software using the \textit{version.bind} in the CHAOS class.
Queries and replies are recorded tied to the \NS that provided them.

We ran these measurements between October 10\(^{th}\) to 14\(^{th}\) and 22\(^{nd}\) to 24\(^{th}\) 2022 against the Alexa Top1M from August 15\(^{th}\) 2022 containing 476,242 zones, collecting responses to a total of 32M queries sent via \ip4 and 24M queries sent via \ip6.
Our active measurement dataset (101GB of json data), and a tool implementing our measurement toolchain are publicly available at: \url{https://github.com/mutax/dns-v6-readyness}

\vspace{-0.5em}
\subsection{Ethical Considerations}
\label{sec:ethics}

The \textit{Farsight Security Information Exchange} (SIE)
dataset~\cite{Farsight-SIE} used in this work is collected by Farsight Inc. at
globally distributed vantage points, co-located to recursive DNS resolvers.
These sensors collect and aggregate DNS cache misses they encounter, i.e.,
outgoing queries of the recursors and the received answers. Only collecting
cache misses is a conscious choice by Farsight to ensure PII is protected. The
dataset also does not contain which sensors collected a specific entry. We
specifically use a per-month aggregated version of the dataset, see
\Cref{sec:method:dataset}. For details on the fields in the dataset, see
\Cref{tab:data}. Data has been handled according to best practices in network
measurement data handling as outlined by Allman and
Paxson~\cite{allman2007issues}.

Before running the active measurements for validation purposes (cf. \Cref{sec:validation}), we consult the Menlo report \cite{kenneally2012menlo} as well as best measurement practices~\cite{durumeric2013zmap}.
We limit our probing rate, send only well-formed DNS requests, and make use of dedicated servers which have informative rDNS names set.
Additionally, we run a webserver providing additional information and contact details on the IP as well as on the rDNS name.
We also focused our measurements on the Alexa Top 1M, i.e., sites for which the impact of additional requests at the scale of our measurements is not significant, while also limiting repeated requests using caching.
During our active measurements, we did not receive any complaints.
In summary, we conclude that this work does not raise any ethical issues.

\section{Results}
\label{sec:results}

Here, we first provide an aggregate overview of the Farsight dataset.
Subsequently, we present the results of our analysis of \broken based on passive measurement data.
Finally, we validate our passive measurement results against active measurements run from 10\(^{th}\) to 24\(^{th}\) of October 2022.

\subsection{Dataset Overview}
\label{sec:results:overview}
\begin{figure}[t!]
	\centering
	\vspace{2em}
	\includegraphics[width=.99\linewidth]{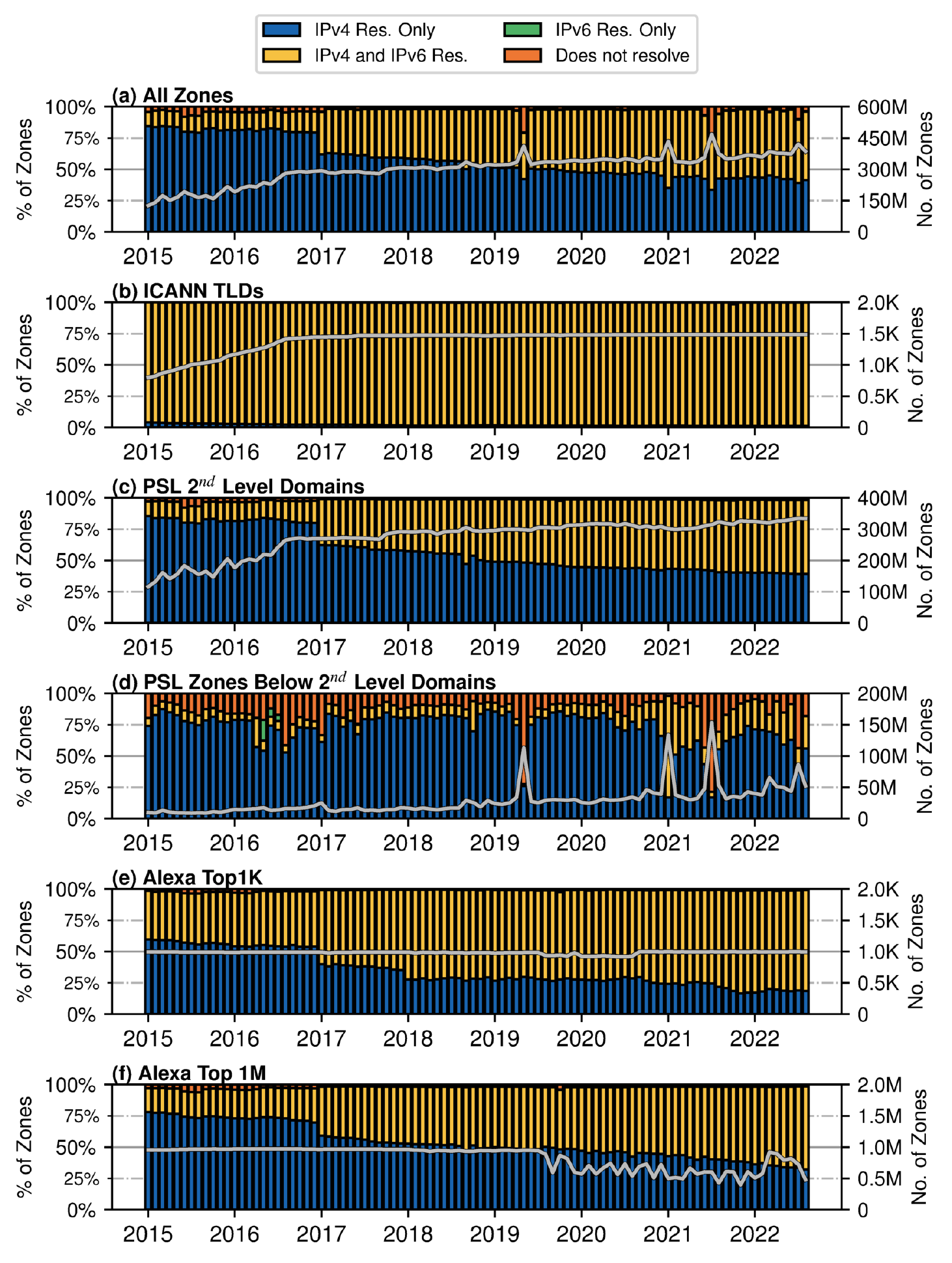}
	\caption{Per month: \# of zones (gray line--right y-axis)
          and IPv4/IPv6 resolvability in \% (left y-axis).}
	\label{fig:zones_summary}
	\vspace{1em}
    \phantomsubfigure{fig:zones_total}
    \phantomsubfigure{fig:zones_icann}
    \phantomsubfigure{fig:zones_psl}
    \phantomsubfigure{fig:zones_psub}
    \phantomsubfigure{fig:zones_alexa1k}
    \phantomsubfigure{fig:zones_alexa}
\end{figure}

Our passive dataset spans 7~years starting on January 1\(^{st}\), 2015 and ending on
\lastdatefull. During this period, the number of unique zones
increased from 126\,M to 368\,M. Similarly, the number of PSL 2\(^{nd}\) level
domains increased from 116\,M to 326\,M. For a visualization see the gray line
in \Cref{fig:zones_summary} (right y-axis). To highlight our
findings, we present results for selected subsets
of domains only. The full results for all domain subsets are in shown in
\Cref{app:figures}.

\subsection{IPv6 Resolution in DNS Over Time}
In \Cref{fig:zones_summary} we show how the fraction of zones that is resolvable via \ipo4,
\ipo6, both protocols, or fails to resolve, changes across time.
We also show how the total number of zones changes
(gray line). The figure shows data for all zones, the ICANN TLDs, PSL 2\(^{nd}\)
domains, zones deeper in the tree, Alexa Top-1K and Alexa Top-1M.

Overall, see \Cref{fig:zones_total}, we find that \vsixresstart of
all zones are \resip6 in January 2015. This is significantly higher than
the sub 1\% reported by Czyz et al.~\cite{czyz} in 2014. However, they
only accounted for glue records, which does not consider zones with out of
bailiwick \NS. Over time \ip6 adoption steadily increases, with \vsixresend of zones
resolving via \ip6 in \lastdate. A notable increase of \ip6 resolvable
zones by \vsixchgjansev occurs in January 2017. Further investigation we find,
that this increase relates to two major DNS providers---a PaaS
provider and a webhoster---adding \AAAA glue for their \NS.

For ICANN TLDs, see \Cref{fig:zones_icann}, we find that the majority of zones
is \resip6.  Throughout our observation period nearly all TLDs are \resip6. The
remaining not \resip6 zones are several vanity TLDs as well as smaller ccTLDs.

While PSL 2\(^{nd}\) level domains, see \Cref{fig:zones_psl},
mirror the general trend of all zones, we find that zones deeper in the tree (\Cref{fig:zones_psub})
are generally less likely to be \resip6. Still, we observe an upward
trend. We attribute this to the fact that the process of entering such domains
into TLDs for 2\(^{nd}\) level domains still receives oversight by NICs, e.g., regarding the RFC compliant
use of at least two \NS in different networks~\cite{rfc2182}, while zones below
2\(^{nd}\) level domains can be freely delegated by their domain owners. 
Also, for sub-domains, we observe three distinct spikes in \Cref{fig:zones_psub} which correspond to the spikes seen for all
domains, recall \Cref{fig:zones_total}. These spikes occur
when a single subtree of the DNS spawns millions of zones. These are
artifacts due to specific configurations and highlight that lower
layer zones may not be representative for the overall state of DNS.

Finally, comparing PSL 2\(^{nd}\) level domains, see \Cref{fig:zones_psub}, to
the Alexa~Top-1K domains, see \Cref{fig:zones_alexa1k}, we find that \ip6
adoption is significantly higher among popular domains, starting from
\alexaonevsixresstart in 2015 and rising to \alexaonevsixresend in 2021. There
are two notable steps in this otherwise gradual increase, namely January 2017
and January 2018. These are due to a major webhoster and a major PaaS provider
enabling \ip6 resolution (2017), and a major search engine provider common in
the Alexa-Top-1K enabling \ip6 resolution (2018).

\vspace{1em}
\noindent\textbf{Comparison with Active Measurements:}
Evaluating zone resolvability from our active measurements, see \Cref{sec:validation}, we find that 314,994 zones (66.14\%) support dual stack DNS resolution, while 159,166 zones (33.42\%) are only resolvable via \ip4.

A further 2066 zones (0.43\%) could not be resolved during our active measurements, and 16 zones (\(\leq\)0.01\%) were only resolvable via \ip6.
In comparison to that, our passive measurements--see also \Cref{fig:zones_alexa}--map closely:
We find 66.18\% (+0.04\% difference) of zones in the Alexa Top 1M resolving via both, \ip4 and \ip6, and 32.23\% (-1.19\% difference) of zones only resolving via \ip4.
Similarly, 1.16\% (+0.73\%) of zones do not resolve at all, and 0.42\% (+0.42\% difference) of zones only resolve via \ip6 according to our passive data.
Hence, overall, we find our passive approach being closely aligned with the results of our active measurements for the latest available samples.
The, in comparison, higher values for non-resolving and \ip6 only resolving zones are most likely rooted in the visibility limitations of the dataset, see \Cref{sec:discussion:limitations}.
Nevertheless, based on the low deviation between two independent approaches at determining \ip6 resolvability of zones we have confidence in the results of our passive measurements.

\subsection{\ip6 Resolution Failure Types}
\begin{figure}[t!]
	\centering
\vspace{4em}
	\includegraphics[width=.99\linewidth]{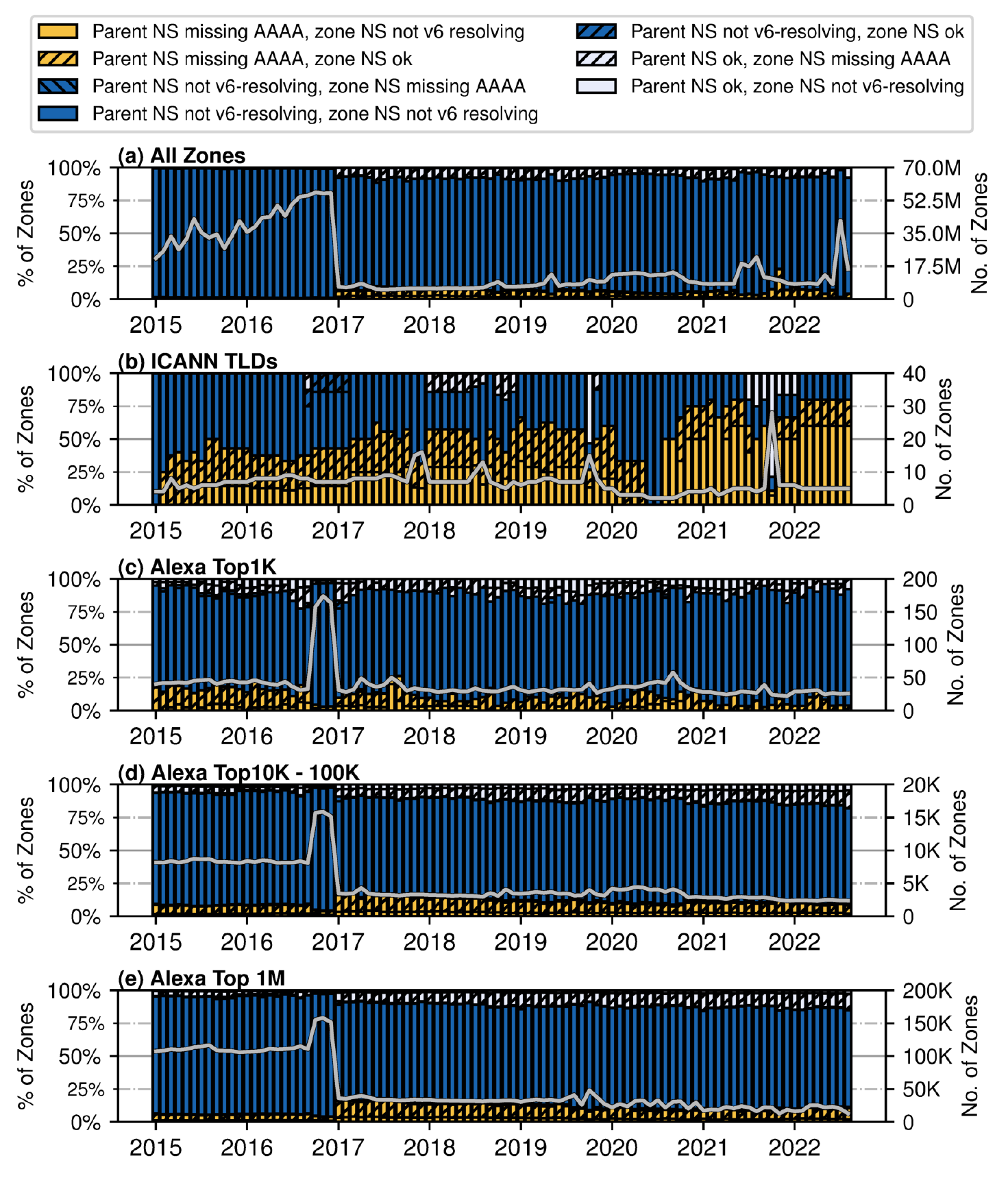}
	\caption{Per month: \# of zones not \resip6
             with \AAAA or \glue for \NS (gray line--right y-axis)
             and causes for \ip6 resolution failure in \% (left y-axis).}
	\label{fig:fail_states}
\vspace{4em}
    \phantomsubfigure{fig:fail_total}
    \phantomsubfigure{fig:fail_icann}
    \phantomsubfigure{fig:fail_alexa1k}
    \phantomsubfigure{fig:fail_alexa100k}
    \phantomsubfigure{fig:fail_alexa}
\end{figure}

Next, we take a closer look at zones that show some indication of \ip6
deployment, yet, are not \resip6. These are zones where an \NS has an
\AAAA record or an \AAAA \GLUE. To find them we consider \NS entries within the
zone as well as \NSes for the zone in its parent. In \Cref{fig:fail_states} we
show how their absolute numbers evolve over time (gray line) as well as the
failures reasons (in percentages).

We find that for all four subsets of zones shown---all zones, ICANN TLDs, Alexa
Top-1K, Alexa Top-10K--100K---the most common failure case is missing
resolution of \NS in the parent.
This occurs mostly when the \NS is out-of-bailiwick and \emph{does} have \AAAA records,
but the \NS's zone itself is not \resip6.
Furthermore, there is a substantial number of zones per category---especially
in the Alexa Top-1K---where the \NS in the parent lacks \AAAA while the \NS
listed in the zone has \AAAA records, commonly due to missing \GLUE. We also
observe the inverse scenario, i.e., \GLUE is present but no \AAAA
record exist for the \NS within the zone itself. Both cases can also occur if
\nssets differ between the parent and its child~\cite{sommese}.

We see a major change around January 2017, i.e., a sharp increase
in zones that are \resip6, which is also visible in \Cref{fig:zones_summary}: For
all zones as well as for the Alexa Top 10K--100K, we observe that several
million zones not resolving via \ip6 since the start of the dataset but
having \NSes with \AAAA records, now are \resip6. The reason is that a major
provider added missing glue records. Interestingly, we do not see this
in the Alexa Top~1K.

In the Alexa Top 1K, and to a lesser degree in the Alexa Top 10K-100K, we
observe a spike of zones that list \AAAA records for their \NS but are not
\ip6-resolvable in Oct.\ 2016. This is the PaaS provider mentioned before,
first rolling out \AAAA records for their \NS, and then three months later also
adding \ip6 \GLUE. Operationally, this approach makes sense, as they can first
test the impact of handling \ip6 DNS queries in general. Moreover, reverting
changes in their own zones is easier than reverting changes in the TLD
zones--here the \GLUE entries. Again, the major webhoster is less common among
the \emph{very} popular domains, which is why its effect can be seen in
\Cref{fig:fail_total,fig:fail_alexa100k}, but not in
\Cref{fig:fail_icann}. Also, this operator had \AAAA records in place since the
beginning of our dataset, as seen by the plateau in
\Cref{fig:fail_alexa100k}. These observations have been cross-confirmed by
inspecting copies of zonefiles for the corresponding TLDs and time-periods.

\subsection{Centralization and \ip6 Readiness}
\label{sec:results:ns}
\begin{figure}[t!]
	\centering
\vspace{4em}
	\includegraphics[width=.99\linewidth]{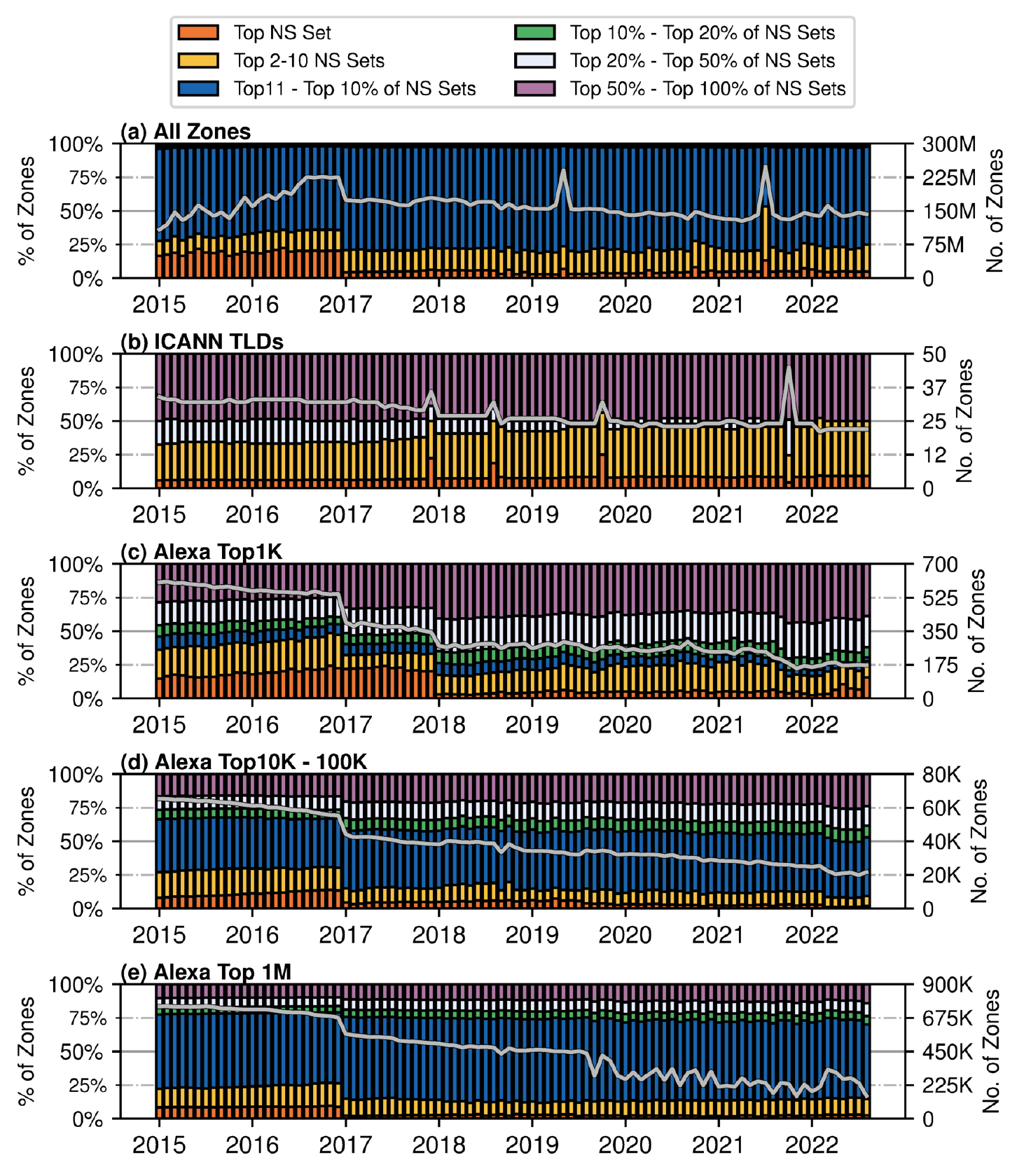}
	\caption{Per month: \# of zones not \resip6 (gray
          line--right y-axis) and distribution of zones over \nssets in \%
          (left y-axis).}
	\label{fig:nssets}
    \phantomsubfigure{fig:nssets_total}
    \phantomsubfigure{fig:nssets_icann}
    \phantomsubfigure{fig:nssets_alexa1k}
    \phantomsubfigure{fig:nssets_alexa100k}
    \phantomsubfigure{fig:nssets_alexa}
\vspace{4em}
\end{figure}

Finally, we focus on the \nss hosting most non \resip6 zones.
We first identify the top \nssets in terms of the number of hosted
zones, aggregating \NS names to their PSL 2\(^{nd}\) level domain and known
operators' \NS under a multiple well-patterned zones. Then, we compute a
CDF over the number of zones per \nsset for
each time bin. \Cref{fig:nssets} shows how this CDF changed across time
and highlights the impact of centralization within the DNS providers.
Over \vsixendtoptenzoneshare of the non-\resip6 zones are hosted by the Top~ \vsixendrespops of \nssets.

Again, we see the impact of a change by a major webhoster in January 2017---it
is the top \nsset among all zones (\Cref{fig:nssets_total}).
Similarly, the PaaS provider is pronounced among the Alexa Top-1K, i.e., part
of the Top 10 of \nssets (\Cref{fig:nssets_alexa1k}) and the top \nsset for the
Alexa Top-10K--100K (\Cref{fig:nssets_alexa100k}). Finally, the major search
engine operator's impact can especially be seen among TLDs
(\Cref{fig:nssets_icann}) and the Alexa Top-1K (\Cref{fig:nssets_alexa1k}),
where---in both cases---this operator is the top \nsset for non \resip6 zones.

\subsection{Resolvability and Responsiveness of \NS in Active Measurements}
During our active measurements, we also had the opportunity to validate whether NS records listed in zones did actually reply to DNS requests or not.
During our evaluation of the Alexa Top 1M, we discovered a total of 176,207 NS records, of which 212 had A or AAAA records associated that were invalid, as for example \verb+::+ as a AAAA record.
Of the remaining 175,995 records, 116,504 needed glue, i.e., they were in-bailiwik NS for their own.
Among these, 19,310 NS were dual-stack, while 94,192 only had A records associated with them, and a further 108 \NS only had associated \AAAA records.
Furthermore, 85,213 (90.47\%) of \A-only \NS needing glue had correct glue set.
For dual-stack configured \NS, 14,072 (72.87\%) have complete (\A and \AAAA) glue.
A further 3,932 (20.36\%) \NS only has \A glue records, while 24 (0.12\%) \NS only have \AAAA glue, despite generally having a dual-stack DNS configuration.
Finally, of the 108 \NS records only having \AAAA records associated, 70 (64.81\%) \NS have correctly set \AAAA glue.

Moving on to the reachability of these \NS, we find that of the total number of \NS that have an \A record (169,547) \emph{and} are reachable is at 164,255, i.e., 96.88\% actually responds to queries.
For \ip6, these values are slightly worse, with 30,193 of 32,285 \NS (93.52\%) responding to queries via \ip6.
This highlights a potential accuracy gap of 3-6\% for research work estimating DNS resolvability from passive data.
Notably, this gap is larger for IPv6.
\vspace{-.5em}
\section{Discussion}
\label{sec:discussion}

In this section, we first state our key-findings, and then discuss their
implications.

\subsection{The Impact of Centralization}
Centralization is one of the big changes in the Internet over the last
decade. This trend ranges from topology
flattening~\cite{bottger2018elusive,arnold2020cloud} to the majority of content being
served by hypergiants~\cite{bottger2017hypergiant} and---as we
show---also applies to the DNS. An increasing number of zones are operated by
a decreasing number of organizations. As such, an outage at one big DNS
provider~\cite{dyn}---or missing support for \ip6---can disrupt name resolution
for a very large part of the Internet as we highlight in
\Cref{sec:results}. In fact, out-of-bailiwick \NS not being resolvable
via \ip6 is the most common misconfiguration in our study, often triggered by
 missing \GLUE in a single zone. Given that \emph{ten} operators could
enable \ip6 DNS resolution for \vsixendtoptenabszoneshare of not yet \ip6 resolving zones, we claim
that large DNS providers have a huge responsibility for
making the Internet IPv6 ready.

\subsection{\ip6 DNS Resolution and the Web}
In general, as we travel down the delegation chain we find more
misconfigurations and a smaller fraction of \ip6-resolvable zones.  Given that
common web assets--JavaScript, Style Sheets, or images--are often served from
FQDNs further down the DNS hierarchy, we conjecture that this may have a
another huge, yet still hidden, impact on IPv6 readiness for web. We encourage
operators to be mindful of this issue, and study its effect in future work.

\subsection{Implications for Future Research}
Our findings demonstrate that it is not sufficient to test for the presence of
\AAAA records to asses the \ip6 readiness of a DNS zone. Instead, measurements
have to assess whether the zones are \ip6-resolvable.  The same applies to
email setups and websites.

Furthermore, given the centralization we observe in the DNS, network
measurements of \ip6 adoption should consider and quantify the impact of
individual operators.  More specifically, researchers should distinguish
between effects caused by a small number of giants vs.\ the behavior of the
Internet at large. Artifacts that can occur temporarily should be
recognized and then excluded.

\subsection{Limitations} \label{sec:discussion:limitations}

Since our dataset relies on DNS cache misses, we are missing domains that are
not requested or not captured by the Farsight monitors in a given month.
Moreover, our use of monthly aggregates may occlude short-term
misconfigurations.  To address this, we support major findings on
misconfigurations with additional ground-truth data from authoritative TLD zone
files. 

Similarly, we use the Alexa List with its known
limitations~\cite{rweyemamu2019clustering,scheitle2018long}.  Thus, we cluster
the Alexa list into different rank tiers, which reduces fluctuations in
the higher tiers.  Furthermore, we only assess zones' configuration states, and not
actual resolution, i.e., ``lame delegation'' for other reasons is out of scope.

Furthermore, we cannot make statements on whether the zones we measure
\emph{actually} resolve, e.g., if there is an authoritative DNS server
listening on a configured IP address returning correct results.  Still, we have
certainty that zones we measure as resolvable are at least sufficiently
configured for resolution.  Similarly, we can not assess the impact of observed
DNS issues on other protocols, e.g., HTTPs.  To further address this limitation
of our passive data source, we conducted active measurements, which validated
the observations from our passive results and added further insights on the
actual reachability of authoritative DNS servers for zones.

Naturally, our active measurements also have several limitations that have to
be recorded.  First, we conducted our measurements from a single vantage point.
Given load balancing in CDNs via DNS~\cite{ednsstreibelt}, this may have lead to a
vantage point specific perspective.  Nevertheless, we argue that
misconfigurations~\cite{dietrich2018investigating} are likely to be consistent across an operator,
i.e., the returned A or AAAA records may change, but not the issue of, e.g.,
missing \GLUE.  Furthermore, DNS infrastructure tends to be less dynamic than
\A and \AAAA records.

Second, our measurements were only limited to the Alexa Top 1M and associated
domains.  We consciously made this choice instead of, e.g., running active
measurements on \emph{all} zones in the Farsight dataset to reduce our impact
on the Internet ecosystem.

In summary, our study provides an important first perspective on IPv6 only
resolvability.  We suggest to complement our study with active measurements of
\ip6 only DNS resolution and the impact of \broken on the \ip6
readiness of the web due to asset dependencies as future work.

\section{Related work}

Our related work broadly clusters into two segments: \emph{i)} Studies on IPv6
adoption and readiness, and \emph{ii)} Studies about DNS and DNS
misconfigurations.

\subsection{IPv6 Adoption and Readiness}
With the exhaustion of the IPv4 address space~\cite{prichterv6}, IPv6 adoption
has been a frequent topic of study. In 2014, Czyz et al.~\cite{czyz} conducted
a primer study on IPv6 adoption, taking a multi-perspective approach that also
covered DNS. Our measurements shed light on the time after their measurements
which concluded in 2014.  Furthermore, they estimate IPv6 adoption in DNS by
only surveying \AAAA glue records in \texttt{net.} and \texttt{com.}, while we
consider the full resolution path.  Work by Foremski et al.~\cite{entropyip}
and Plonka \& Berger~\cite{temporalv6} investigate IPv6 adoption at the edge,
which is orthogonal to our work. In recent years, various researchers took
country and domain specific perspectives on IPv6 adoption,
e.g.,~\cite{v6_china, v6_africa, pamv6_2010}.

\subsection{DNS and DNS Misconfiguration Studies}
Since DNS is a core component of the Internet, it has been studied regularly
over the past decades, including studies regarding the adoption of new protocol
features,
e.g.,~\cite{doan,dnssec17,ednscalder,ednsstreibelt,dnshttps,dnstlspam}.  Such
studies use various active datasets, e.g., OpenINTEL~\cite{openintel}, as well
as passive datasets, e.g., the Farsight SIE dataset which we rely on, to, e.g.,
study operational aspects of the DNS~\cite{foremski2019dns}. More specifically
focusing on DNS (mis)configuration, Sommese et al.~\cite{sommese} study
inconsistencies in parent and child \nssets and Akiwate et al.~\cite{akiwate}
work on lame delegation.
However, contrary to our work, the latter two either do not consider the IP
part of DNS delegation (Sommese et al.), or explicitly focus on IPv4 (Akiwate et
al.).
More recently, Izhikevich et al. presented ZDNS, a tool for large-scale studies of the DNS ecosystem in the Internet~\cite{izhikevich2022zdns}.
Unfortunately, ZDNS is tailored towards IPv4 and does not support querying authoritative nameservers over IPv6.
Therefore, we cannot make use of ZDNS in our study.
Instead we perform active DNS measurements with our own implementation of a DNS resolution methodology, which implements \ip6 resolution.

\subsection{Summary} We expand on earlier contributions regarding IPv6
adoption. We provide a more recent perspective on the IPv6 DNS ecosystem and
take a more complete approach to asses the IPv6 readiness in an \ipo6 scenario.
This focus on IPv6 is also our novelty in context to earlier work on DNS
measurements and DNS misconfigurations, which did not focus on how IPv6 affects
DNS resolvability. Additionally, our active measurements for validating our
passive measurement results also highlight that the presence of \AAAA records
does not necessarily imply \ip6 resolvability.  Instead, to measure \ip6
resolvability, the resolution state of provided \ip6 resources has to be
validated.
\section{Conclusion}
\vspace{0,5em}
In this paper, we present a passive DNS measurement study on root causes for
\broken in an \ip6 only setting. While over time we see an increasing number of
zones resolvable via \ip4 and \ip6, in \lastdate still \lastunresvsix are not
resolvable via \ip6. We identify not resolvable \NS records of the zone or its
parent as the most common failure scenario. Our recommendations to operators
include to explicitly monitor \ip6 across the entire delegation chain. 
\vspace{0,5em}

Additionally, we conducted a dedicated validation of our results using active 
measurements. This validation broadly confirmed our results from the passive
measurements and further highlighted the importance of not only relying on the
presence of specific records, as nameservers for which \ip6 addresses are listed
in the DNS may not actually be responsive.

\vspace{0,5em}
We plan to provide an open-source implementation of our measurement methodology
along with the paper.  Furthermore, we will provide a reduced implementation of
our measurement toolchain which will enable operators to explicitly check a
given zone or FQDN for \resip6. Similarly, we will provide the results of our
active measurements as open data.

\vspace{0,5em}
For future work we suggest to systematically expand our active measurement
campaign to assess resolvability, \eg for websites including all web assets.
Using active measurements, one can explicitly resolve a hostname and run active
checks on the delegation chain, validating the responses of all \anss and find
inconsistencies not only between a zone and its parent but also within the
\nsset.  We conjecture that--especially given the widespread use of subdomains
for web assets--the reduced \ip6 resolvability we observe may have a
significant impact on the \ip6-readiness of the web, i.e., a website using
assets on domains that do not resolve via \ip6 is not \ip6 ready.
\section*{Acknowledgments}
We thank Farsight Security, Inc. (now DomainTools) for providing access to the
Farsight Security Information Exchange's passive DNS data feed.  Without this
data, the project would not have been possible.  The authors express their
gratitude to the anonymous reviewers for their thoughtful and encouraging input
during the reviewing process.  manner.  This work was partially funded by the
German Federal Ministry of Education and Research under the project PRIMEnet,
grant 16KIS1370, and 6G-RIC, grant 16KISK027.  Any opinions, findings, and
conclusions or recommendations expressed in this material are those of the
authors and do not necessarily reflect the views of Farsight Security, Inc.,
DomainTools, the German Federal Ministry of Education and Research or the
authors' host institutions and further affiliations.

\newpage
\appendix

\section{DNS Resolution Overview}
\label{app:figures}

\vspace{-1em}

\begin{figure}[H]
        \centering
        \includegraphics[width=.7\linewidth]{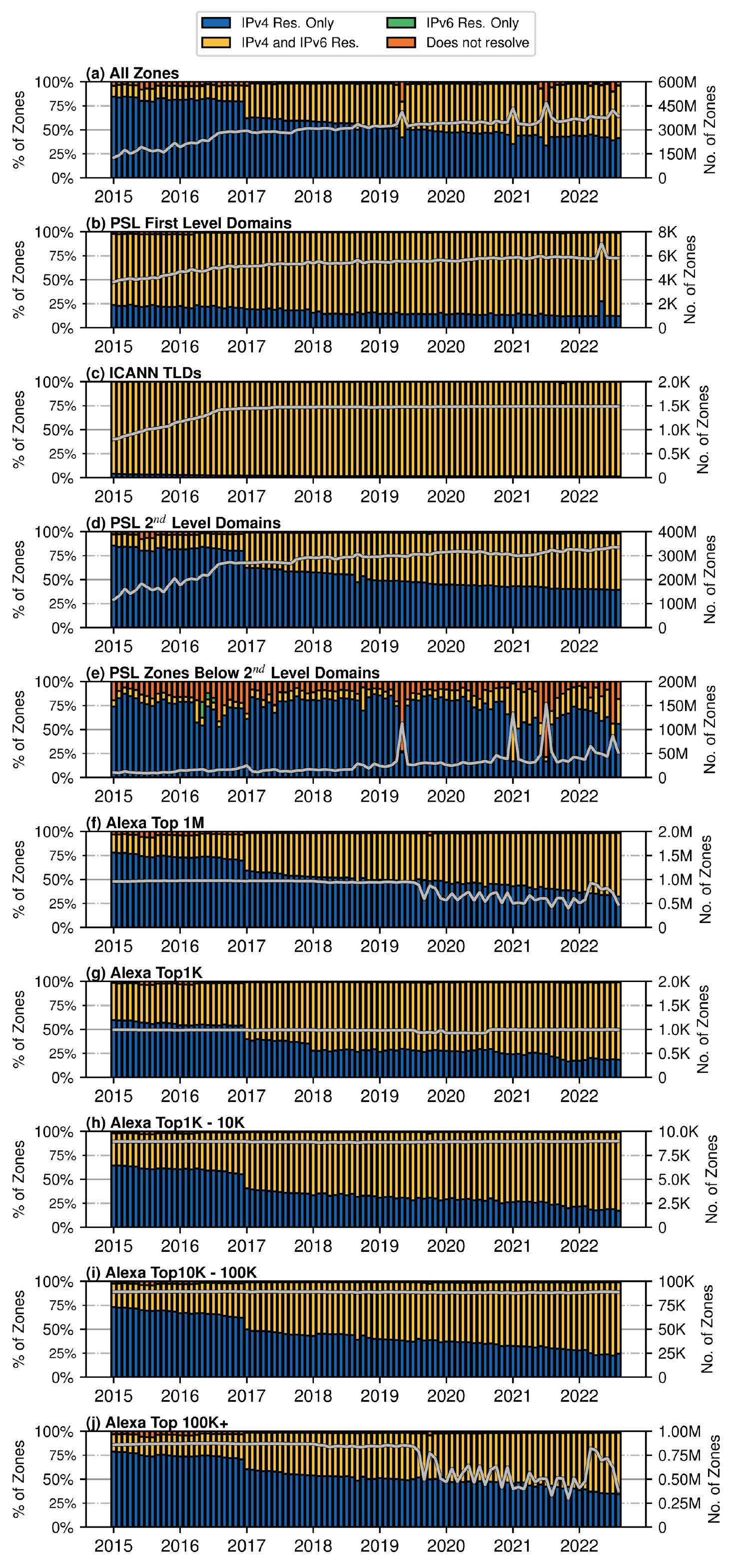}
        \caption{Total number of zones in the dataset per month (gray line) and resolvability}
        \label{fig:apdx_resolution_summary}
    \phantomsubfigure{fig:apdx_zone_total}
    \phantomsubfigure{fig:apdx_zone_psl}
    \phantomsubfigure{fig:apdx_zone_icann}
    \phantomsubfigure{fig:apdx_zone_psld}
    \phantomsubfigure{fig:apdx_zone_psub}
\end{figure}

\section{IPv6 Only Resolution Failures}
\vspace{-1em}
\begin{figure}[H]
        \centering
        \includegraphics[width=.7\linewidth]{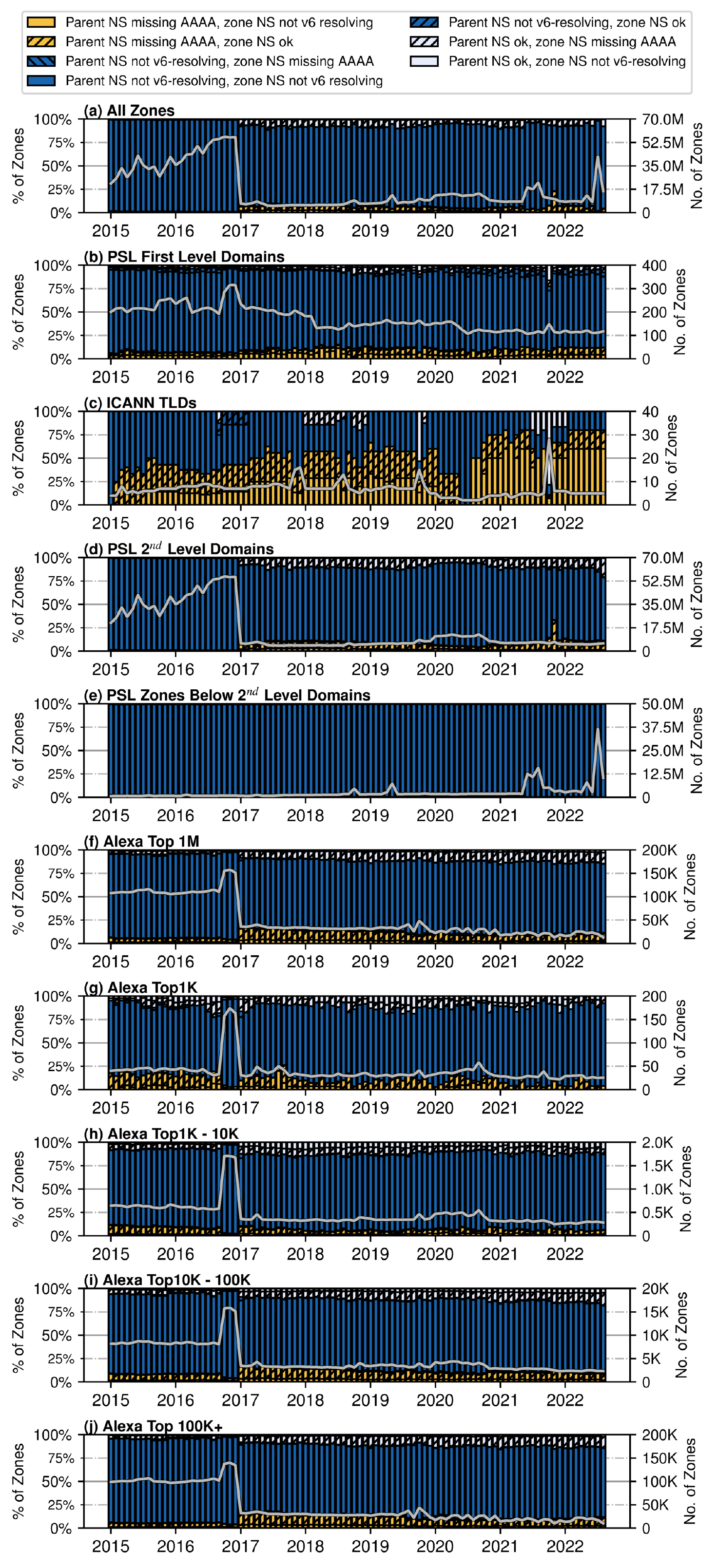}
        \caption{Zones unable to resolve using \ip6,
             but with \AAAA records in \glue or zone apex (gray line),
             by resolution failure.}
        \label{fig:apdx_fail_states}
    \phantomsubfigure{fig:apdx_fail_total}
    \phantomsubfigure{fig:apdx_fail_psl}
    \phantomsubfigure{fig:apdx_fail_icann}
    \phantomsubfigure{fig:apdx_fail_psld}
    \phantomsubfigure{fig:apdx_fail_psub}
\end{figure}

\section{Zones Without IPv6 Resolution per \nsset}
\vspace{-1em}
\begin{figure}[H]
    \vspace{9pt}
        \centering
        \includegraphics[width=.7\linewidth]{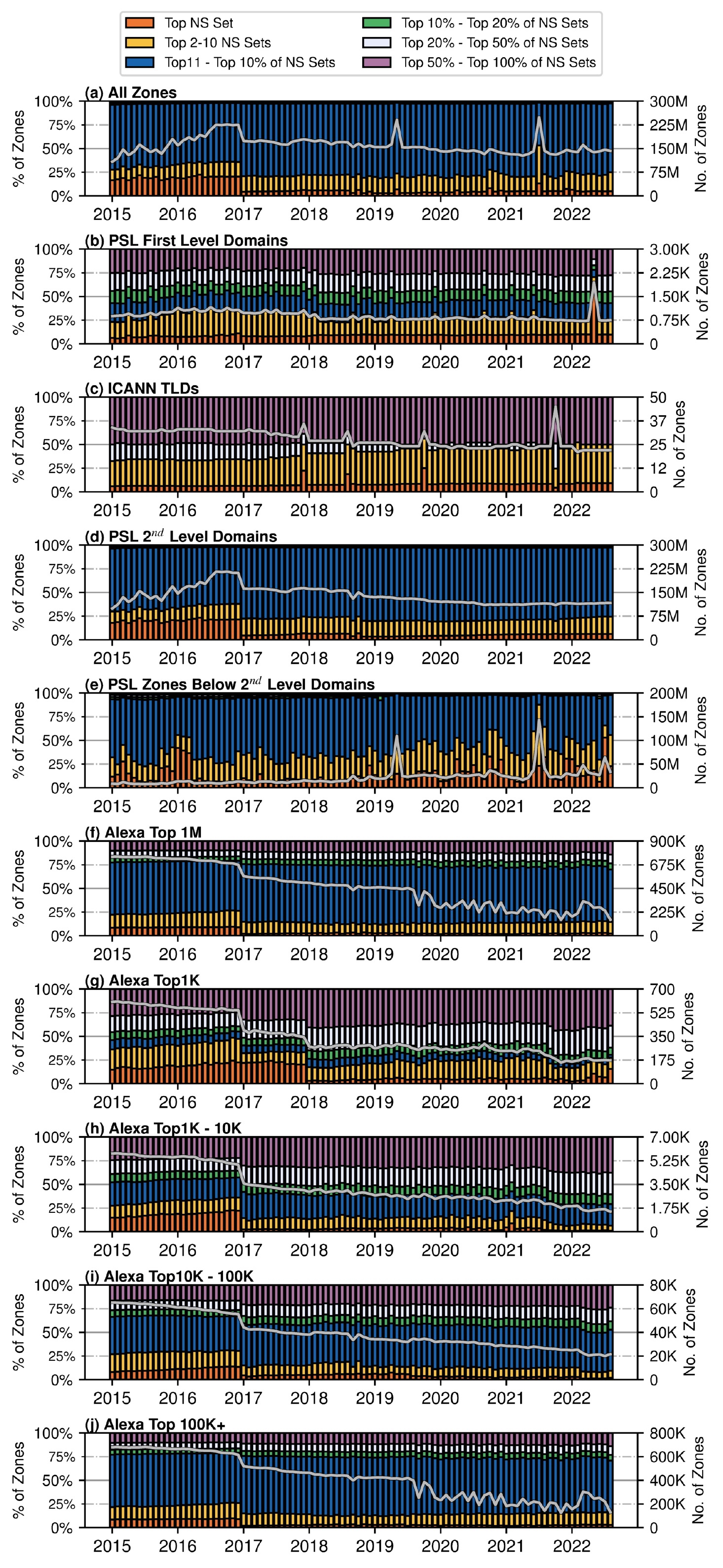}
        \caption{Distribution of zones not resolving via \ip6 over \nssets.}
        \label{fig:apdx_nssets}
    \phantomsubfigure{fig:apdx_nsset_total}
    \phantomsubfigure{fig:apdx_nsset_psl}
    \phantomsubfigure{fig:apdx_nsset_icann}
    \phantomsubfigure{fig:apdx_nsset_psld}
    \phantomsubfigure{fig:apdx_nsset_psub}
\end{figure}


\begin{thebibliography}{10}
\providecommand{\url}[1]{\texttt{#1}}
\providecommand{\urlprefix}{URL }
\providecommand{\doi}[1]{https://doi.org/#1}

\bibitem{akiwate}
Akiwate, G., Jonker, M., Sommese, R., Foster, I., Voelker, G.M., Savage, S.,
  Claffy, K.: {Unresolved Issues: Prevalence, Persistence, and Perils of Lame
  Delegations}. In: Proc. of the Internet Measurement Conference (IMC). p.
  281–294. ACM (2020). \doi{10.1145/3419394.3423623}

\bibitem{allman2007issues}
Allman, M., Paxson, V.: {Issues and Etiquette Concerning Use of Shared
  Measurement Data}. In: Proc. of the Internet Measurement Conference (IMC).
  pp. 135--140. ACM (2007). \doi{10.1145/1298306.1298327}

\bibitem{Alexa}
{Amazon.com, Inc}: {Alexa Top Sites}. \url{https://www.alexa.com/}

\bibitem{arnold2020cloud}
Arnold, T., He, J., Jiang, W., Calder, M., Cunha, I., Giotsas, V.,
  Katz-Bassett, E.: {Cloud provider connectivity in the flat Internet}. In:
  Proc. of the Internet Measurement Conference (IMC). pp. 230--246. ACM (2020).
  \doi{10.1145/3419394.3423613}

\bibitem{rfc7816}
Bortzmeyer, S.: {DNS Query Name Minimisation to Improve Privacy}. RFC 7816
  (Experimental) (Mar 2016), \url{https://www.rfc-editor.org/rfc/rfc7816.txt},
  obsoleted by RFC 9156

\bibitem{rfc8020}
Bortzmeyer, S., Huque, S.: {NXDOMAIN: There Really Is Nothing Underneath}. RFC
  8020 (Proposed Standard) (Nov 2016),
  \url{https://www.rfc-editor.org/rfc/rfc8020.txt}

\bibitem{bottger2018elusive}
B{\"o}ttger, T., Antichi, G., Fernandes, E.L., di~Lallo, R., Bruyere, M.,
  Uhlig, S., Castro, I.: {Shaping the Internet: 10 Years of IXP Growth}. arXiv
  (2019). \doi{10.48550/ARXIV.1810.10963},
  \url{https://arxiv.org/abs/1810.10963}

\bibitem{bottger2017hypergiant}
B{\"o}ttger, T., Cuadrado, F., Tyson, G., Castro, I., Uhlig, S.: {A
  Hypergiant’s View of the Internet}. ACM Computer Communication Review (CCR)
   \textbf{47}(1) (2017)

\bibitem{ednscalder}
Calder, M., Fan, X., Hu, Z., Katz-Bassett, E., Heidemann, J., Govindan, R.:
  {Mapping the Expansion of Google's Serving Infrastructure}. In: Proc. of the
  Internet Measurement Conference (IMC). p. 313–326. ACM (2013).
  \doi{10.1145/2504730.2504754}

\bibitem{dnshttps}
Chhabra, R., Murley, P., Kumar, D., Bailey, M., Wang, G.: {Measuring
  DNS-over-HTTPS Performance around the World}. In: Proc. of the Internet
  Measurement Conference (IMC). p. 351–365. ACM (2021).
  \doi{10.1145/3487552.3487849}

\bibitem{dnssec17}
Chung, T., van Rijswijk-Deij, R., Choffnes, D., Levin, D., Maggs, B.M.,
  Mislove, A., Wilson, C.: {Understanding the Role of Registrars in DNSSEC
  Deployment}. In: Proc. of the Internet Measurement Conference (IMC). p.
  369–383. ACM (2017). \doi{10.1145/3131365.3131373}

\bibitem{pamv6_2010}
Colitti, L., Gunderson, S.H., Kline, E., Refice, T.: {Evaluating IPv6 Adoption
  in the Internet}. In: Proc. of the Passive and Active Measurement Conference
  (PAM). pp. 141--150. Lecture Notes in Computer Science, Springer (2010)

\bibitem{czyz}
Czyz, J., Allman, M., Zhang, J., Iekel-Johnson, S., Osterweil, E., Bailey, M.:
  {Measuring IPv6 Adoption}. In: Proc. of the 2014 ACM SIGCOMM Conference
  (SIGCOMM). p. 87–98. ACM (2014). \doi{10.1145/2619239.2626295}

\bibitem{dietrich2018investigating}
Dietrich, C., Krombholz, K., Borgolte, K., Fiebig, T.: {Investigating system
  operators' perspective on security misconfigurations}. In: Proc. of the 25th
  ACM SIGSAC Conference on Computer and Communications Security (CCS). pp.
  1272--1289. ACM (2018)

\bibitem{doan}
Doan, T.V., Fries, J., Bajpai, V.: {Evaluating Public DNS Services in the Wake
  of Increasing Centralization of DNS}. In: IFIP Networking Conference (2021).
  \doi{10.23919/IFIPNetworking52078.2021.9472831}

\bibitem{dnstlspam}
Doan, T.V., Tsareva, I., Bajpai, V.: {Measuring DNS over TLS from the Edge:
  Adoption, Reliability, and Response Times}. In: Hohlfeld, O., Lutu, A.,
  Levin, D. (eds.) Proc. of the Passive and Active Measurement Conference
  (PAM). pp. 192--209. Lecture Notes in Computer Science, Springer (2021)

\bibitem{Farsight-SIE}
{DomainTools, formerly Farsight Security}: {Farsight Security Information
  Exchange (SIE)}.
  \url{https://www.farsightsecurity.com/solutions/security-information-exchange/}
  (2022)

\bibitem{rfc3901}
Durand, A., Ihren, J.: {DNS IPv6 Transport Operational Guidelines}. RFC 3901
  (Best Current Practice) (Sep 2004),
  \url{https://www.rfc-editor.org/rfc/rfc3901.txt}

\bibitem{durumeric2013zmap}
Durumeric, Z., Wustrow, E., Halderman, J.A.: {ZMap: Fast Internet-wide Scanning
  and Its Security Applications}. In: Proc. of the 31th USENIX Security
  Symposium (USENIX Security). pp. 605--620. USENIX Association (2022)

\bibitem{rfc2181}
Elz, R., Bush, R.: {Clarifications to the DNS Specification}. RFC 2181
  (Proposed Standard) (Jul 1997),
  \url{https://www.rfc-editor.org/rfc/rfc2181.txt}, updated by RFCs 4035, 2535,
  4343, 4033, 4034, 5452, 8767

\bibitem{rfc2182}
Elz, R., Bush, R., Bradner, S., Patton, M.: {Selection and Operation of
  Secondary DNS Servers}. RFC 2182 (Best Current Practice) (Jul 1997),
  \url{https://www.rfc-editor.org/rfc/rfc2182.txt}

\bibitem{fiebig2017something}
Fiebig, T., Borgolte, K., Hao, S., Kruegel, C., Vigna, G.: {Something from
  nothing (There): collecting global IPv6 datasets from DNS}. In: Proc. of the
  Passive and Active Measurement Conference (PAM). Lecture Notes in Computer
  Science, vol. 10176, pp. 30--43. Springer (2017)

\bibitem{foremski2019dns}
Foremski, P., Gasser, O., Moura, G.C.: {DNS observatory: The big picture of the
  DNS}. In: Proc. of the Internet Measurement Conference (IMC). pp. 87--100.
  ACM (2019)

\bibitem{entropyip}
Foremski, P., Plonka, D., Berger, A.: {Entropy/IP: Uncovering Structure in IPv6
  Addresses}. In: Proc. of the Internet Measurement Conference (IMC). p.
  167–181. ACM (2016). \doi{10.1145/2987443.2987445}

\bibitem{v6_china}
Han, C., Li, Z., Xie, G., Uhlig, S., Wu, Y., Li, L., Ge, J., Liu, Y.: {Insights
  into the issue in IPv6 adoption: A view from the Chinese IPv6 Application
  mix}. Concurrency and Computation: Practice and Experience  \textbf{28}(3),
  616--630 (2016). \doi{https://doi.org/10.1002/cpe.3327}

\bibitem{rfc8499}
Hoffman, P., Sullivan, A., Fujiwara, K.: {DNS Terminology}. RFC 8499 (Best
  Current Practice) (Jan 2019),
  \url{https://www.rfc-editor.org/rfc/rfc8499.txt}

\bibitem{houser2021comprehensive}
Houser, R., Hao, S., Li, Z., Liu, D., Cotton, C., Wang, H.: {A Comprehensive
  Measurement-based Investigation of DNS Hijacking}. In: Proc. of the 40th
  International Symp. on Reliable Distributed Systems (SRDS). pp. 210--221.
  IEEE (2021)

\bibitem{tlds}
{ICANN}: {List of Top-Level Domains}.
  \url{https://www.icann.org/resources/pages/tlds-2012-02-25-en}

\bibitem{izhikevich2022zdns}
Izhikevich, L., Akiwate, G., Berger, B., Drakontaidis, S., Ascheman, A.,
  Pearce, P., Adrian, D., Durumeric, Z.: {ZDNS: A Fast DNS Toolkit for Internet
  Measurement}. In: Proc. of the Internet Measurement Conference (IMC). ACM
  (2022)

\bibitem{kenneally2012menlo}
Kenneally, E., Dittrich, D.: {The Menlo Report: Ethical Principles Guiding
  Information and Communication Technology Research}. Available at SSRN 2445102
   (2012)

\bibitem{le2019tranco}
Le~Pochat, V., Van~Goethem, T., Tajalizadehkhoob, S., Joosen, W.: {TRANCO: A
  Research-Oriented Top Sites Ranking Hardened Against Manipulation}. In: Proc.
  of the 26th Network and Distributed System Security Symposium (NDSS).
  Internet Society (ISOC) (2019)

\bibitem{liu2018reexamination}
Liu, B., Lu, C., Li, Z., Liu, Y., Duan, H.X., Hao, S., Zhang, Z.: {A
  Reexamination of Internationalized Domain Names: The Good, the Bad and the
  Ugly.} In: Proc. of the 48th IEEE/IFIP International Conference on Dependable
  Systems and Networks (DSN). pp. 654--665. IEEE (2018)

\bibitem{v6_africa}
Livadariu, I., Elmokashfi, A., Dhamdhere, A.: {Measuring IPv6 Adoption in
  Africa}. In: e-Infrastructure and e-Services for Developing Countries. pp.
  345--351. Springer International Publishing, Cham (2018)

\bibitem{psl}
{Mozilla Foundation}: {Public Suffix List}. \url{https://publicsuffix.org/}

\bibitem{unbound}
{NLnet Labs}: {Unbound Nameserver Documentation}.
  \url{https://unbound.docs.nlnetlabs.nl/en/latest/reference/history/requirements.html}

\bibitem{openintel}
{OpenINTEL project}: {The OpenINTEL measurement platform}.
  \url{https://openintel.nl/}

\bibitem{temporalv6}
Plonka, D., Berger, A.: {Temporal and Spatial Classification of Active IPv6
  Addresses}. In: Proc. of the Internet Measurement Conference (IMC). p.
  509–522. ACM (2015). \doi{10.1145/2815675.2815678}

\bibitem{prichterv6}
Richter, P., Allman, M., Bush, R., Paxson, V.: {A Primer on IPv4 Scarcity}. ACM
  Computer Communication Review (CCR)  \textbf{45}(2),  21–31 (2015).
  \doi{10.1145/2766330.2766335}

\bibitem{rweyemamu2019clustering}
Rweyemamu, W., Lauinger, T., Wilson, C., Robertson, W., Kirda, E.: {Clustering
  and the weekend effect: Recommendations for the use of top domain lists in
  security research}. In: Proc. of the Passive and Active Measurement
  Conference (PAM). pp. 161--177. Lecture Notes in Computer Science, Springer
  (2019)

\bibitem{scheitle2018long}
Scheitle, Q., Hohlfeld, O., Gamba, J., Jelten, J., Zimmermann, T., Strowes,
  S.D., Vallina-Rodriguez, N.: {A long way to the top: Significance, structure,
  and stability of Internet top lists}. In: Proc. of the Internet Measurement
  Conference (IMC). pp. 478--493. ACM (2018)

\bibitem{rfc8305}
Schinazi, D., Pauly, T.: {Happy Eyeballs Version 2: Better Connectivity Using
  Concurrency}. RFC 8305 (Proposed Standard) (Dec 2017),
  \url{https://www.rfc-editor.org/rfc/rfc8305.txt}

\bibitem{sommese}
Sommese, R., Moura, G.C.M., Jonker, M., van Rijswijk-Deij, R., Dainotti, A.,
  Claffy, K.C., Sperotto, A.: {When Parents and Children Disagree: Diving into
  DNS Delegation Inconsistency}. In: Sperotto, A., Dainotti, A., Stiller, B.
  (eds.) Proc. of the 15th Passive and Active Measurement Conference (PAM). pp.
  175--189. Lecture Notes in Computer Science, Springer (2020)

\bibitem{ednsstreibelt}
Streibelt, F., B\"{o}ttger, J., Chatzis, N., Smaragdakis, G., Feldmann, A.:
  {Exploring EDNS-Client-Subnet Adopters in Your Free Time}. In: Proc. of the
  Internet Measurement Conference (IMC). p. 305–312. ACM (2013).
  \doi{10.1145/2504730.2504767}

\bibitem{dyn}
{ThousandEyes Blog, Cisco}: {The DDoS Attack on Dyn’s DNS Infrastructure}.
  \url{https://www.thousandeyes.com/blog/dyn-dns-ddos-attack/}

\bibitem{rfc6555}
Wing, D., Yourtchenko, A.: {Happy Eyeballs: Success with Dual-Stack Hosts}. RFC
  6555 (Proposed Standard) (Apr 2012),
  \url{https://www.rfc-editor.org/rfc/rfc6555.txt}, obsoleted by RFC 8305

\end{thebibliography}
\end{document}